\documentclass[usenatbib]{mn2e}
\newcommand{\bib}{\bibitem[\protect\citeauthoryear}
\usepackage{epsf}
\usepackage[T1]{fontenc}
\usepackage{ae,aecompl}
\begin{document}
\title[3C31 adiabatic jet model]{Adiabatic relativistic models for the jets
in the radio galaxy 3C\,31}
\author[R.A. Laing \& A.H. Bridle]{R.A. Laing
       \thanks{E-mail: rlaing@eso.org}$^{1,2}$, A.H. Bridle$^3$\\
     $^1$ European Southern Observatory,
     Karl-Schwarzschild-Stra\ss e 2, D-85748 Garching-bei-M\"{u}nchen,
     Germany \\
     $^2$ Astrophysics, University of Oxford, Denys
     Wilkinson Building, Keble Road, Oxford OX1 3RH \\
     $^3$ National Radio Astronomy Observatory, 520 Edgemont Road,
     Charlottesville, VA 22903-2475, U.S.A.}

\date{Received }
\maketitle

\begin{abstract}
We present a general approach to the modelling of the brightness and
polarization structures of adiabatic, decelerating relativistic jets,
based on the formalism of \citet{MS}.  We compare the predictions of
adiabatic jet models with deep, high-resolution observations of the radio
jets in the FR\,I radio galaxy 3C\,31.  Adiabatic models require coupling
between the variations of velocity, magnetic field and particle density.
They are therefore more tightly constrained than the models previously
presented for 3C\,31 by \citet{LB02a}. We show that adiabatic models
provide a poorer description of the data in two crucial respects: they
cannot reproduce the observed magnetic-field structures in detail, and
they also predict too steep a brightness decline along the jets for
plausible variations of the jet velocity.  We find that the innermost
regions of the jets show the strongest evidence for non-adiabatic
behaviour, and that the adiabatic models provide progressively better
descriptions of the jet emission at larger distances from the galactic
nucleus.  We briefly discuss physical processes which might contribute to
this non-adiabatic behaviour. In particular, we develop a parameterized
description of distributed particle injection, which we fit to the
observed total intensities.  We show that particles are preferentially
injected where bright X-ray emission is observed, and where we infer that
the jets are over-pressured.
\end{abstract}

\begin{keywords}
galaxies: jets -- radio continuum:galaxies -- magnetic fields --
polarization -- MHD -- acceleration of particles
\end{keywords}

\section{Introduction}
\label{Introduction}

This paper is one of a series whose aim is to develop quantitative models of
jets in low-luminosity (FR\,I; \citealt{FR74}) radio galaxies on the assumption
that they are decelerating relativistic flows. Our hypothesis is that the jets
are close enough to being intrinsically symmetrical, axisymmetric and
antiparallel that the observed differences between them are dominated by the
effects of relativistic aberration.  This hypothesis is motivated by the results
of \citet{LPdRF}, who presented a statistical study of jets in the B2 sample of
radio galaxies and showed that the observed correlations between fractional core
flux density and side-to-side asymmetries in intensity and width are consistent
with jet deceleration and the presence of transverse velocity gradients. They
inferred that the jets slow from $\approx$0.9$c$ where they start to expand
rapidly to $\approx$0.1$c$ over distances $\sim$1 to 10\,kpc and that the
deceleration scale is an increasing function of jet power.  In
\citet[][hereafter LB]{LB02a}, we demonstrated that an intrinsically
symmetrical, relativistic jet model provides an excellent description of the
total intensity and linear polarization observed at 8.4\,GHz from the jets of
the FR\,I radio galaxy 3C\,31.  We were able to estimate the angle to the line of
sight and the three-dimensional distributions of velocity, emissivity and
magnetic-field structure.  By combining this kinematic model with a description
of the external gas density and pressure derived from {\em Chandra} observations
\citep{Hard02} and using conservation of particles, energy and momentum, we
demonstrated that the jet deceleration could be produced by entrainment of
thermal matter and we derived the spatial variations of pressure, density and
entrainment rate \citep{LB02b}.

The {\em free models} developed by LB for 3C\,31 were designed to fit the
observed images without embodying specific preconceptions about the (poorly
known) internal physics. In those models, we adopted simple and arbitrary
functional forms for the spatial variations of velocity, synchrotron emissivity
and field ordering and allowed the emissivity and field ordering to vary
separately as smooth functions of position.  To proceed beyond such purely
empirical descriptions of the jets, we must make further assumptions about
processes that affect the components of the modelled emissivity -- the
relativistic particle energy spectrum and the strength of the magnetic field.
The separation of the emissivity into its components is ill-determined unless
inverse-Compton emission can be detected from the synchrotron-emitting
electrons, in which case the particle density and field strength can be
determined independently. Unfortunately, inverse-Compton emission from the inner
jets in 3C\,31 is too weak to detect \citep{Hard02}.  There are, as yet, no
prescriptive theories for dissipative processes such as particle acceleration
and field amplification or reconnection in conditions appropriate to FR\,I radio
jets.

The energy-loss processes for the radiating particles can be quantified,
however.  It is inevitable that the particles will suffer adiabatic losses as
the jets expand.  We argue that synchrotron and inverse-Compton losses are
negligible by comparison for electrons radiating at 8.4\,GHz in 3C\,31, as the
jets have accurately power-law spectra with indices close to 0.5
\citep{LPBFGMP04} and show no evidence of spectral curvature until much higher
frequencies \citep{Hard02}.  It is therefore worthwhile to compare the
observations with models in which the radiating particles are accelerated before
entering the region of interest and then lose energy only by the adiabatic
mechanism while the magnetic field is frozen into and convected passively with
the flow, which is assumed to be laminar.  Following conventional usage, we
refer to such models as {\em adiabatic}.

There are analytical solutions for the adiabatic evolution of emissivity
and surface brightness with jet radius and velocity if the magnetic field
is exactly parallel or perpendicular to the flow direction.  Models of
this kind were first considered by \citet{Bur79}, who showed that the
decrease of brightness with distance from the nucleus would be much
steeper than he observed in 3C\,31 if the jets have constant
velocity. \citet{Fan82} pointed out that the brightness would decline more
slowly with distance in a decelerating jet. This idea underlies the
turbulent jet models of \citet{ Bic84,Bic86}.  A relativistic
generalisation was developed by \citet{Bau97} to model the jets in 3C\,264
and was applied by \citet {Fer99} and \citet{ Bondi00} to other FR\,I
jets.  These treatments all assumed that there is no velocity gradient
across the jet, and that the magnetic field is exactly parallel or
perpendicular to the flow. While this approach is self-consistent, it
cannot be applied if there is velocity shear in a direction perpendicular
to any component of the field.

In LB, we developed a numerical approach to modelling a relativistic jet
with a more general magnetic field structure.  This approach allowed us to
use the variations of total intensity and linear polarization as
independent constraints on the jet velocity field.  We concluded that the
magnetic structure and velocity field in 3C\,31 are indeed significantly
more complex than those assumed by \citet{Bau97}. Nevertheless, we
compared our models and the data for 3C\,31 with their analytical
solutions.  We showed that:

\begin{enumerate}
\item the adiabatic approximation is qualitatively inconsistent with the
variations of brightness and polarization along the first $\approx$3 kpc
of the jets, but
\item further from the nucleus, the observed variations are closer to
those expected if the adiabatic approximation holds.
\end{enumerate}
This comparison motivated us to develop a more general approach to
modelling of adiabatic, relativistic jets, which we present in this paper.  Our
approach allows us to calculate how the brightness and polarization
structure evolve along an adiabatic jet from {\it prescribed initial
conditions} (specified as profiles across the jet), given the jet geometry
and more complex magnetic structures and velocity fields of the type
inferred for 3C\,31 by LB.  Detailed comparison of the new adiabatic
models with the free models of LB, which fit the observations better, can
then diagnose whether and where other physical processes, such
as particle acceleration, may be significant in the jets.

Section~\ref{ad-approx} reviews our assumptions and the previous analytical
solutions, and then outlines our calculation of synchrotron emission from
adiabatic flows. Section~\ref{Obs-mod} describes our approach to modelling of
adiabatic jets; it briefly recapitulates material from LB before discussing new
aspects specific to the present study. Section~\ref{Outer-region} applies our
adiabatic models to the outer regions of the jets in 3C\,31 and shows that they
can give a fair description of the VLA observations of these
regions. Section~\ref{Whole-jet} confirms that the adiabatic models fail to
describe the inner jet regions and critiques the adiabatic hypothesis in the
light of this result; it also discusses the extent to which distributed particle
injection can bring the adiabatic models into better agreement with the
data. Section~\ref{Conclusions} summarizes our conclusions.

\section{The adiabatic approximation}
\label{ad-approx}

\subsection{Assumptions}
\label{assumptions}

The jets are taken to be adiabatic in the sense defined by \citet{Bur79}, \citet
{MS} and \citet{Bau97}, as follows:
\begin{enumerate}
\item The energies of the radiating particles change like those in an
adiabatically expanding relativistic gas, i.e.\ $\propto V^{-1/3}$, where
$V$ is the volume of a fluid element in its rest frame.
\item There is no diffusion of particles.
\item The particle momentum distribution remains isotropic (e.g.\ by
resonant scattering off Alfv\'{e}n waves).
\item The magnetic field behaves as if it is convected passively with the
flow. We take the velocity field to be a smooth function of position.
Although the addition of a turbulent velocity component would make little
difference to the calculation of the effects of relativistic aberration on
the appearance of the jet, there would be a major difference in the
strength and structure of the magnetic field, which would be affected by
shear and expansion, if not by dynamo action and reconnection.
\item Synchrotron and inverse-Compton losses are negligible for electrons
radiating at the wavelength of observation, so the energy and synchrotron
frequency spectra are always power laws with indices $-(2\alpha+1)$ and
$-\alpha$, respectively.  Specifically, we take the number density of
radiating electrons with energies between $E$ and $E + dE$ in the jet rest
frame to be
\begin{equation}
N(E) dE = n E^{-(2\alpha+1)} dE \label{eq-energy-spec}
\end{equation}
\end{enumerate}
In addition, we assume, as in LB, that the regions of the jets
to be modelled are intrinsically identical, antiparallel, axisymmetric,
stationary flows.

The magnetic field $B$ is taken to have longitudinal, toroidal and
radial components $B_l$, $B_t$ and $B_r$ (all measured in the jet rest
frame).  Synchrotron radiation is generally anisotropic even in the
rest frame of the emitting material. In this paper we therefore write
the emissivity $\epsilon({\bf r}) g({\bf r})$, where $\epsilon({\bf
r})$ would be the emissivity in total intensity for a magnetic field
$\langle B_l^2 + B_t^2 + B_r^2 \rangle^{1/2}$ perpendicular to the line
of sight ($\langle\rangle$ denotes a spatial average). $\epsilon({\bf
r})$ is the same for all three Stokes parameters, but $g({\bf r})$
depends on field geometry, and differs for $I$, $Q$ and $U$ [for total
intensity, $0 \leq g_I({\bf r}) \leq 1$ and for linear polarization
$-p_0 \leq g_{Q, U}({\bf r}) \leq p_0$, where $p_0 =
(3\alpha+3)/(3\alpha+5)$ is the maximum degree of polarization for
spectral index $\alpha$].

\subsection{Analytical approximations}
\label{analytic}

We now briefly recapitulate the analytical adiabatic relations derived
by \citet{Bau97} for  axisymmetric, decelerating,
relativistic jets without velocity shear.  Suppose that a jet has
radius $R$, and that we can make the quasi-one-dimensional approximation:
(a) that it is uniform in cross-section at a given distance from the
nucleus and (b) that the velocity is unidirectional.  The field components and
the normalizing constant in the energy spectrum, $n$, vary as:
\begin{eqnarray}
B_l & \propto & R^{-2} \label{eq-Bl}\\
B_t & \propto & (R\beta\Gamma)^{-1} \label{eq-Bt}\\
B_r & \propto & (R\beta\Gamma)^{-1} \label{eq-Br}\\
n & \propto & (\Gamma\beta R^2)^{-(1 + 2\alpha/3)} \label{n0-anal} \label{eq-n0}\\
\nonumber
\end{eqnarray}
where the velocity of the jet is $\beta c$ and $\Gamma =
(1-\beta^2)^{-1/2}$.

For a purely longitudinal field ($B_t = B_r = 0$), this leads
to a variation of the rest-frame emissivity:
\begin{eqnarray}
\epsilon & \propto & (\Gamma\beta)^{-(2\alpha+3)/3}R^{-(10\alpha+12)/3} \\
\nonumber \end{eqnarray}
and for a perpendicular field ($B_l = 0$):
\begin{eqnarray}
\epsilon & \propto & (\Gamma\beta)^{-(5\alpha+6)/3}R^{-(7 \alpha+ 9)/3}
\label{em-prof-eqn}\\ \nonumber\end{eqnarray} For conical jets, the radius
is proportional to distance from the nucleus, $r$, so identical relations
are obtained with $r$ replacing $R$.

In the absence of velocity shear, the adiabatic relations for the magnetic
field can be combined to describe arbitrary initial conditions. The total
field $B$ is then:
\begin{equation}
B = \left[ \bar{B}_l^2\left(\frac{\bar{R}}{R}\right)^4
+ (\bar{B}_t^2 +
\bar{B}_r^2)\left(\frac{\bar{\Gamma}\bar{\beta}\bar{R}}{\Gamma\beta
R}\right)^2\right ]^{1/2}
\end{equation}
where $\bar{B}_l$, $\bar{B}_t$ and $\bar{B}_r$ are the initial field
components, $\bar{\beta}$ and $\bar{\Gamma}$ are the velocity and Lorentz
factor, all defined where the jet radius is $R = \bar{R}$.  The emissivity
is then:
\begin{equation}
\epsilon \propto (\Gamma\beta R^2)^{-(1 + 2\alpha/3)} B^{1+\alpha}
\end{equation}
The next
subsection develops a numerical approach capable of describing velocity
shear.

\subsection{Adiabatic flows with arbitrary initial conditions and velocity
fields}
\label{numeric}

The formalism needed to predict the total and linearly polarized synchrotron
emission from an element of fluid in a non-relativistic adiabatic flow was
first developed by \citet{MS}. Their approach was to follow an element of
fluid containing relativistic particles and an initially isotropic,
disordered magnetic field through a model flow.  They included the effects
of synchrotron and inverse-Compton energy losses, as well as adiabatic
effects. \citet{L02} developed a simplified approach for the
 case where synchrotron and inverse-Compton losses are negligible, which
we follow here. The reasons for taking the field to be disordered on small
scales are discussed by \citet{L81}, \citet*{BBR} and LB.

An element of fluid is taken to be a unit cube (much smaller than the
spatial scale of variations in the velocity field), with sides defined by
orthonormal unit vectors ${\bf \hat{a}}$, ${\bf \hat{b}}$ and ${\bf
\hat{c}}$ along natural axes of the flow in the fluid rest frame.  The
cube contains an isotropic disordered field
 and relativistic particles with the energy spectrum of
equation~(\ref{eq-energy-spec}) and normalizing constant $\bar{n}$.  The
cube moves with the flow, deforming into a parallelepiped with sides
defined by the vectors ${\bf a}$, ${\bf b}$ and ${\bf c}$. Application of
flux conservation shows that an element of field which is initially ${\bf
\bar{B}} = \bar{B}_x {\bf \hat{a}} + \bar{B}_y {\bf \hat{b}} + \bar{B}_z
{\bf \hat{c}}$ becomes:
\begin{equation}
{\bf B} = \frac{\bar{B}_x {\bf a} + \bar{B}_y {\bf b} + \bar{B}_z
{\bf c}}{V}
\end{equation}
where $V = {\bf a \cdot b \times c}$ is the volume of the parallelepiped
evaluated in the fluid rest frame.  The energy spectrum constant $n$
evolves according to:
\begin{equation}
n  =  \bar{n} V^{-(1 + 2\alpha/3)} \label{n-eqn}
\end{equation}

We assume that the flow is axisymmetric, with deceleration and a
transverse velocity gradient. We define a second coordinate system
$(x,y,z)$, again in the observed frame, with $z$ along the jet axis, $x$
perpendicular to it in a plane containing the line of sight and $y$ in the
plane of the sky.  We parameterize the streamline by an index $s$, which
varies from 0 at the inner edge of a flow component to 1 at its outer
edge, as in LB. Without loss of generality, we can consider flow
in the $x,z$ plane, where the distance of a streamline from the jet axis is
$x(z,s)$.

\begin{figure}
\epsfxsize=8.5cm
\epsffile{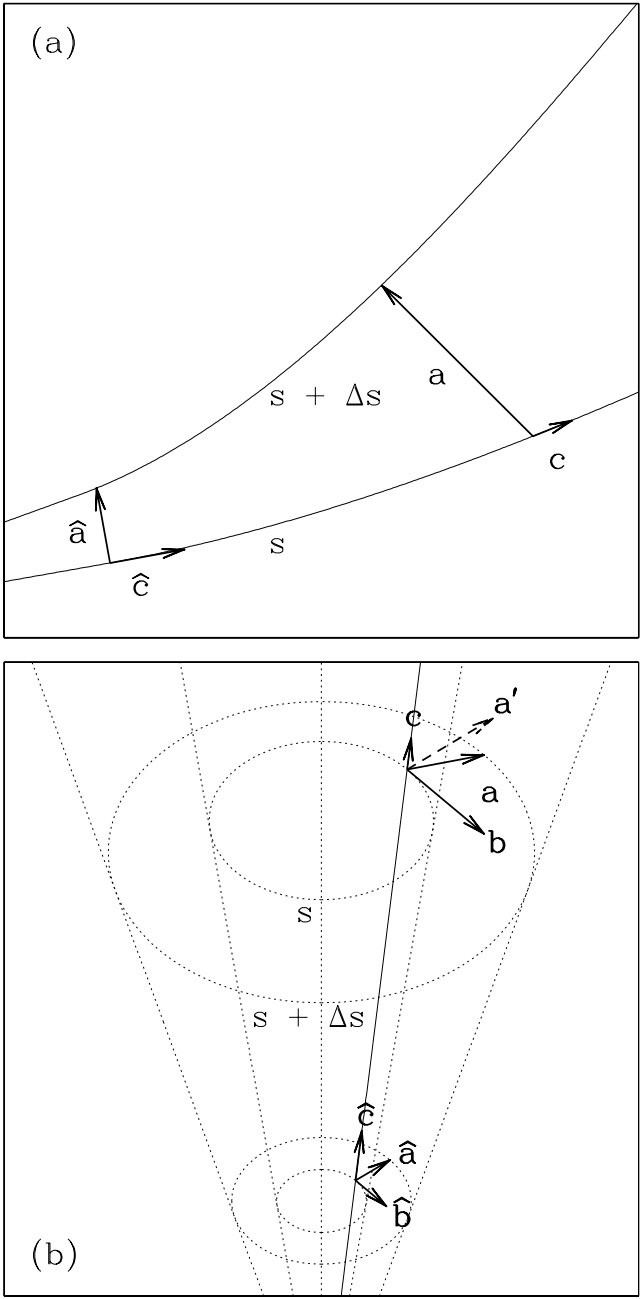}
\caption{Sketches of the evolution of the vectors ${\bf a}$, ${\bf b}$ and
${\bf c}$ defined in the text (not to scale).  Panel (a) shows two
neighbouring streamlines with indices $s$ and $s + \Delta s$ (defined in
Section~\ref{numeric}), in a plane containing the jet axis in a part of the
jet where the flow direction changes with distance from the nucleus. ${\bf
\hat{b}}$ and ${\bf b}$ are perpendicular to the plane of the diagram, and
are not shown.  Note that ${\bf c}$ remains parallel to the flow despite
the change in direction.  Panel (b) shows a projection of the flow in
three dimensions viewed with the jet axis at 45$^\circ$ to the line of
sight in a region where the streamlines are straight.  The full line is a
representative streamline.  The ellipses represent the positions of fluid
elements with streamline indices $s$ and $s + \Delta s$. The lower pair
are at the surface where the initial conditions are set and the upper pair
represent the positions reached by the fluid elements after a given time.
The vector ${\bf a^\prime}$, shown dashed, is the value of ${\bf a}$ in
the absence of shear.
\label{abc-evol}}
\end{figure}

The geometry and evolution of the vectors ${\bf a}$, ${\bf b}$ and ${\bf
c}$ are sketched in Fig.~\ref{abc-evol}.  As an element of fluid moves
outwards, ${\bf c}$ remains parallel to the streamline and its magnitude
is determined entirely by the flow velocity.  ${\bf b}$ is unaffected by
shear and is orthogonal to the flow direction, so its magnitude is
proportional to the distance of the streamline from the jet axis.  ${\bf
a}$ (initially radial) is the only one of the three vectors affected by
velocity shear: it is a function of the jet radius and the accumulated
path length difference along adjacent streamlines
(Fig.~\ref{abc-evol}). Four quantities are therefore needed to describe
the shape of the parallelepiped: three expansion factors and a shear
term. For a streamline in the $xz$ plane, the vectors are:
\begin{eqnarray}
{\bf a} & = & \frac{\partial x /\partial s}{\partial \bar{x}/\partial
s}\left(\frac{1+\bar{x}^{\prime 2}}{1+x^{\prime 2}}\right)^{1/2}{\bf l} +
\frac{\Gamma}{\bar{\Gamma}} f{\bf n} \label{a-eqn}\\
{\bf b} & = & \frac{x}{\bar{x}}{\bf m}\label{b-eqn} \\
{\bf c} & = &
\frac{\Gamma \beta}{\bar{\Gamma}\bar{\beta}}{\bf n} \label{c-eqn}\\ \nonumber
\end{eqnarray}
and the volume is
\begin{equation}
V = {\bf a \cdot b \times c} =
\frac{\Gamma\beta}{\bar{\Gamma}\bar{\beta}}
\frac{x}{\bar{x}} \frac{\partial x /\partial s}{\partial \bar{x}/\partial
s}\left(\frac{1+\bar{x}^{\prime 2}}{1+x^{\prime 2}}\right)^{1/2} \label{vol-eqn}
\end{equation}
${\bf l}$, ${\bf m}$ and ${\bf n}$ are (orthonormal) unit vectors along the local
radial, toroidal and longitudinal directions, respectively and primes
denote differentiation with respect to $z$ for a given streamline. Barred
quantities are evaluated at the starting surface.  The factor
of $\Gamma/\bar{\Gamma}$ in equations~\ref{a-eqn} and \ref{c-eqn} accounts
for Lorentz contraction along the flow direction.

$f$ is the shear term, which does not affect the volume.  It will usually
be $<0$ if the velocity decreases outwards from the jet axis. We evaluate
it by considering two elements of fluid which leave the reference surface
at time $t = 0$ on adjacent streamlines with indices $s$ and $s + \Delta
s$.  After a time $t$, we have
\begin{eqnarray}
ct & = & \int \frac{dl}{\beta} \nonumber \\
   & = & \int^{z(s)}_{\bar{z}(s)}
\frac{[1+x^\prime(z,s)^2]^{1/2} dz}{\beta(z,s)} \nonumber \\
& = & \int^{z(s+\Delta s)}_{\bar{z}(s+\Delta s)}
\frac{[1+x^\prime(z,s+\Delta s)^2]^{1/2} dz}{\beta(z,s+\Delta s)} \nonumber \\
\end{eqnarray}
where $dl$ is an element of path along the streamline and $\bar{z}(s)$ and
$z(s)$ are the $z$-coordinates of the fluid element at the reference
surface and after time $t$, respectively, We express the integral for
streamline $s + \Delta s$ as the sum of the integral for streamline $s$
and a set of terms expanded to first order in $\Delta s$ and $\Delta z =
z(s+\Delta s) - z(s)$. The reference surface is spherical, and is always
set in a part of the jet where the streamlines are straight, so a
streamline crosses the surface at $z = \bar{z}$ with
\begin{equation}
x(\bar{z},s) = \bar{r}\cos[\bar{\zeta}+(\bar{\xi}-\bar{\zeta})s]
\end{equation}
(see Section~\ref{Geom}). This allows us to evaluate $\Delta z/\Delta s$ using
the relation
\begin{eqnarray}
\frac{\Delta z}{\Delta s} \left[ \frac{1+x^{\prime 2}}{\beta} \right] &=&
-\frac{(\bar{\xi}-\bar{\zeta})\bar{r}\tan[\bar{\zeta}+(\bar{\xi}-\bar{\zeta})s]}
{\bar{\beta}} \nonumber \\
&-& \int^{z}_{\bar{z}}\frac{\beta x^\prime\partial x^\prime/\partial s
- (1+x^{\prime 2})\partial\beta/\partial
s}{\beta^2(1+x^{\prime 2})^{1/2}} dz \nonumber \\ \label{eq-dzs}
\end{eqnarray}
The difference in path length along the streamline, $\Delta l$, is given by
\begin{equation}
\frac{\Delta l}{\Delta s} = (1+x^{\prime 2})^{1/2}\left[\frac{\Delta
z}{\Delta s} + \frac{\partial x}{\partial s}\frac{x^\prime}{1+x^{\prime
2}} \right ] \label{eq-dls}
\end{equation}
Finally, the shear term $f$ is the component of ${\bf a}$ along the
streamline normalized by the magnitude of the vectors at the reference
surface:
\begin{equation}
f = \frac{1}{\bar{r}(\bar{\xi}-\bar{\zeta})}\frac{\Delta l}{\Delta s}
\label{eq-shearfactor}
\end{equation}
$f$ is negative if the velocity decreases with increasing $s$.

In general, calculation of the shear term requires a numerical
integration, but for regions of the jet where the streamlines are straight
and the variation of the velocity along a streamline is a simple function,
it can be done analytically.  Note also that the shear term is non-zero
even for a velocity independent of $s$ if the streamlines are curved.

For flow radially outwards from the nucleus, the three vectors and
the volume element can be written very simply in terms of the distance
from the nucleus, $r = (x^2+y^2+z^2)^{1/2}$:
\begin{eqnarray}
{\bf a} & = & \frac{r}{\bar{r}} {\bf \hat{a}} +
\frac{\Gamma}{\bar{\Gamma}} f{\bf \hat{c}}\label{a-eqn-lin} \\
{\bf b} & =
& \frac{r}{\bar{r}} {\bf \hat{b}}\label{b-eqn-lin} \\
 {\bf c} & = &
\frac{\Gamma\beta}{\bar{\Gamma}\bar{\beta}} {\bf \hat{c}}\label{c-eqn-lin}
\\
V & = & \frac{\Gamma \beta r^2}{\bar{\Gamma}\bar{\beta}
\bar{r}^2}\\
\nonumber
\end{eqnarray}
In the absence of shear ($f = 0$) these are equivalent to the relations
derived by \citet{Bau97} and given in equations~\ref{eq-Bl}--\ref{eq-n0}.

\citet{L02} gives the variation of the field components perpendicular to
the line of sight in terms of the direction cosines of the vectors ${\bf
a}$, ${\bf b}$ and ${\bf c}$ with respect to a fixed coordinate system
$(X,Y,Z)$ with $Z$ along the line of sight. Together with $n$, these
determine the synchrotron emissivity in Stokes $I$, $Q$ and $U$.

There are two additional complications:
\begin{enumerate}
\item The direction cosines must be evaluated in the rest frame of the
fluid (exactly as for the free models of LB).
\item We need to set initial conditions on a reference surface in the jet,
rather than starting with an isotropic magnetic field.  We do this by
adjusting the relative magnitudes of the vectors ${\bf \hat{a}}$, ${\bf
\hat{b}}$ and ${\bf \hat{c}}$ in such a way as to produce the desired
ratios between the three field components whilst leaving the volume (and
therefore the rms total field and the emissivity function, $\epsilon$)
unchanged.
\end{enumerate}

\section{Observations and modelling methods}
\label{Obs-mod}

\subsection{Observations}
\label{Obs}

\begin{figure}
\epsfxsize=8.5cm
\epsffile{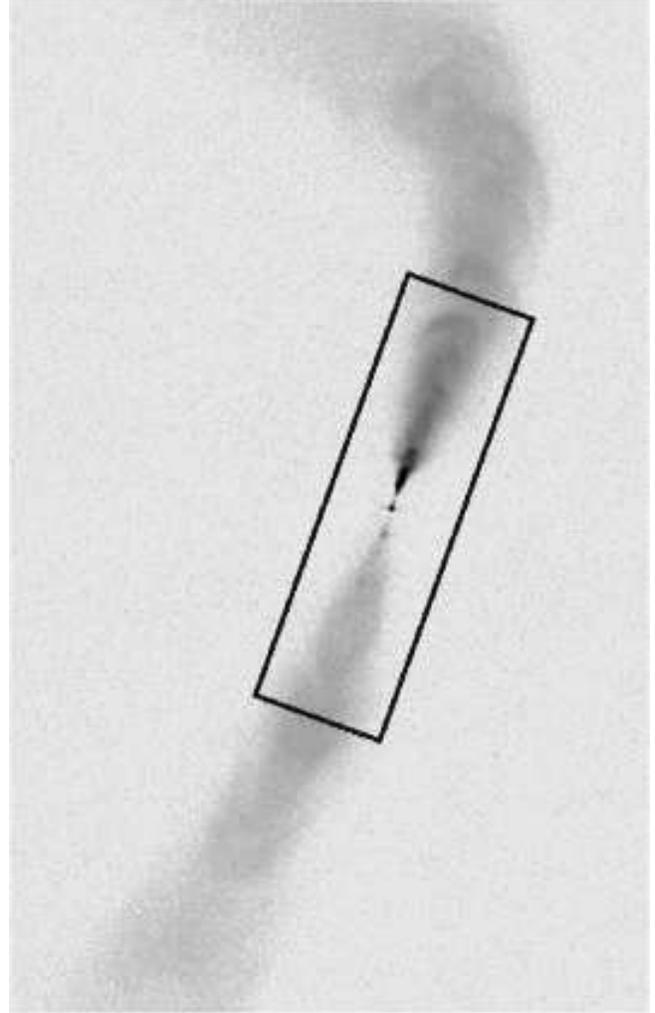}
\caption{A grey-scale of the total intensity from the jets of 3C\,31 at
8.4\,GHz (LB). The resolution is 0.25\,arcsec FWHM and the area we
model is indicated by the box.\label{I0.25}}
\end{figure}

The deep, high-resolution VLA observations with which we compare our
models are described in LB.  We take the Hubble constant to be
$H_0$ = 70\,km s$^{-1}$\,Mpc$^{-1}$.  At the redshift of 3C\,31 (0.0169;
\citealt{Smith2000}, \citealt*{HVG}, \citealt{RC3}), the linear scale
is then 0.34\,kpc/arcsec.  We fit our models to images at resolutions of 0.75 and
0.25\,arcsec, covering the inner $\pm$28\,arcsec of the
jets. Fig.~\ref{I0.25} shows the emission from the jets of 3C\,31 with the
area we model outlined.

\subsection{Geometry}
\label{Geom}

In LB, we showed that the jets in 3C\,31 could be divided into
three regions according to the shapes of their outer isophotes.  As
observed (i.e.\ projected on the plane of the sky), these are:
\begin{enumerate}
\item {\it Inner (0 -- 2.5\,arcsec):} a cone, centred on the nucleus, with a
projected half-opening angle of 8.5 degrees.
\item {\it Flaring (2.5 -- 8.3\,arcsec):} a region in which the jet initially
expands much more rapidly and then recollimates.
\item {\it Outer (8.3 -- 28.3\,arcsec):} a second region of conical
expansion, also centred on the nucleus, but with a projected half-opening
angle of 16.75 degrees.
\end{enumerate}
We showed that these regions also have distinct kinematic
properties.

We use the same descriptions of jet geometry and velocity as in LB, where
we investigated two different transverse velocity structures.  In the
first ({\em spine/shear layer -- SSL}) a central fast spine with no
transverse variation of velocity is surrounded by a slower shear layer
with a linear velocity gradient. In the second ({\em Gaussian}), there is
no distinct spine component, and the jet consists entirely of a shear
layer with a truncated Gaussian velocity law.  The streamline index, $s$
(defined in table\,3 of LB) varies from 0 for the streamline closest to
the axis in the spine or shear layer to 1 for the furthest streamline.
The angles $\bar{\zeta}$ and $\bar{\xi}$ used in Section~\ref{numeric} are
the opening angles of the spine and shear layer, respectively, at the
reference surface for an adiabatic model.

\begin{table*}
\caption{The functional variation of velocity along the model
jets.\label{Long-funcs}}
\begin{tabular}{lllllll}
\hline
&&&&&&\\
Quantity &\multicolumn{4}{c}{Free parameters}
         &\multicolumn{2}{c}{Functional dependences} \\
         & $r_1$  & $r_0$    & $r_{\rm f}$    & Other    & \multicolumn{1}{c}{Flaring} &
 \multicolumn{1}{c}{Outer} \\
&&&&&&\\
\hline
&&&&&&\\
&\multicolumn{5}{c}{On-axis velocities}\\
$\beta_\rho(\rho)$&$\beta_1$&$\beta_0$&$\beta_{\rm f}$&$H$& $b_0 + b_1
         \rho^{H-1} + b_2 \rho^H$ &$c_0 \exp(-c_1 \rho)$\\
&&&&&&\\
&\multicolumn{5}{c}{Fractional edge velocities}\\
$\bar{v}(\rho)$ & $v_1$ & $v_0$ & & &   $v_1 +
	 \frac{(\rho-r_1)(v_0-v_1)}{r_0-r_1}$
& $v_0$ \\
&&&&&&\\
\hline
\end{tabular}
\end{table*}

\begin{figure}
\epsfxsize=8.5cm
\epsffile{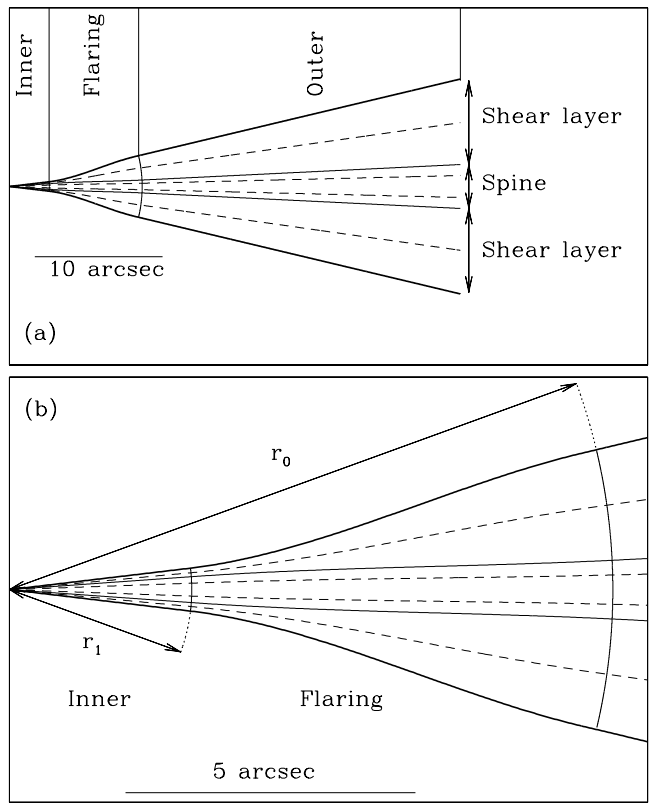}
\caption{Geometry of the spine/shear-layer model, showing the inner,
flaring and outer regions in the plane containing the jet axis. The thick
full curves represent the edge of the jet, the boundaries between regions
are represented by thin full curves and the $s = 0.5$ streamlines for the
spine and shear layer are drawn as dashed curves. (a) The entire modelled
region; (b) the base of the jet on a larger scale, showing the boundary
surfaces at distances of $r_1$ and $r_0$ from the nucleus. The Gaussian
model is essentially the same, but with the spine component removed.
\label{Geom-sketch}}
\end{figure}

The angle between the jet axis and the line of sight is taken to be
$\theta$ and we use the $xyz$ coordinate system defined earlier. In the
inner and outer regions the streamlines are straight. In the flaring
region we interpolate using a cubic polynomial in such a way that $x(z,s)$
and its first spatial derivative $x^\prime(z,s)$ are continuous at the
boundaries between regions, which are spherical, centred on the nucleus
with radii $r_1$ and $r_0$.  The assumed geometry is sketched in
Fig.~\ref{Geom-sketch}.

In order to describe velocity variations along a streamline, we use a coordinate
$\rho$, defined as:
\begin{eqnarray*}
\rho & = & r  \makebox{~~(inner region)} \\
\rho & = & r_1 + (r_0-r_1)\frac{z - r_1\cos \phi_{\rm i}(s)}{r_0 \cos \phi_{\rm o}(s)
- r_1 \cos \phi_{\rm i}(s)}\\
&&  \makebox{~~~~~(flaring region)} \\
\rho & = & r \makebox{~~(outer region)} \\
\end{eqnarray*}
where the streamline makes angles $\phi_{\rm i}$ and $\phi_{\rm o}$ with
the axis in the inner and outer regions, respectively.  $\rho$ is
monotonic along any streamline and varies smoothly from $r_1$ to $r_0$
through the flaring region. This allows us to match on to simple
velocity profiles which depend only on $r$ in the inner and outer regions.

The functions defining the edge of the jet are constrained to match the
observed outer isophotes and are fixed in a coordinate system projected on
the sky.  Their values in the jet coordinate system then depend only on
the angle to the line of sight.

\subsection{Velocity field}

The velocity field is taken to be a separable function $\beta(\rho, s)
= \beta_\rho(\rho) \beta_s(s)$ with $\beta_s(0) = 1$.  The variation
along a streamline, $\beta_\rho(\rho)$ is given in
Table~\ref{Long-funcs}, and is exactly as in LB, omitting the inner
region, which we do not discuss quantitatively in this paper. It is
defined by the index $H$, together with velocities at three locations
in the jet: $r_1$, $r_0$ and an arbitrary fiducial distance
$r_f$. These distances are fixed by their projections on the plane of
the sky, which are 2.5\,arcsec, 8.2\,arcsec and 22.4\,arcsec,
respectively.
The form of $\beta_\rho(\rho)$ was chosen by LB (Section 3.4) to permit
fitting the sidedness-ratio profile observed in 3C\,31, which requires
the velocity to remain constant through most of the flaring region but
then to drop abruptly close to the outer boundary, at a rate determined
by the index $H$. The velocity then falls smoothly and slowly through
the outer region.

$\beta_s(s) = 1$ in the spine for SSL models, dropping linearly with $s$
from 1 at the spine/shear layer interface to a minimum value at the edge
of the jet.  For Gaussian models, $\beta_s(s)$ is a truncated Gaussian
function.  In all models, the minimum fractional velocity,
$\bar{v}(\rho)$, is allowed to vary along the jet
(Table~\ref{Long-funcs}).

\begin{table}
\caption{Summary of the functional variations of velocity, emissivity,
and field component ratios across the jet.\label{Trans-funcs}}
\begin{tabular}{ll}
\hline
&\\
Location & \multicolumn{1}{c}{Functional variation}\\
&\\
\hline
&\\
\multicolumn{2}{c}{Transverse velocity profile (varies along jet)}\\
Spine & $\beta_s(s) = 1$ \\
Shear layer SSL &  $\beta_s(s) = 1 + [\bar{v}(\rho)-1]s$\\
Shear layer Gaussian & $\beta_s(s) = \exp [-s^2 \ln \bar{v}(\rho)]$ \\
&\\
\multicolumn{2}{c}{Emissivity profile at $\rho = r = \bar{r}$}\\
Spine       &  $\epsilon(\bar{r},s) = \bar{E}$ \\
Shear layer SSL & $\epsilon(\bar{r},s) = 1 + (\bar{e}-1)s$ \\
Shear layer Gaussian&  $\epsilon(\bar{r},s) = \exp (s^2 \ln \bar{e})$\\
&\\
\multicolumn{2}{c}{Radial/toroidal field ratio at $\rho = r = \bar{r}$}\\
Spine & $j(\bar{r},s) = \bar{j}_{\rm spine}$\\
Shear layer & $j(\bar{r},s) = \bar{j}_{\rm cen} +
s^p(\bar{j}_{\rm out} - \bar{j}_{\rm cen})$\\
&\\
\multicolumn{2}{c}{Longitudinal/toroidal field ratio at $\rho = r = \bar{r}$}\\
Spine & $k(\bar{r},s) = \bar{k}_{\rm spine}$\\
Shear layer & $k(\bar{r},s) = \bar{k}_{\rm cen} +
s^q(\bar{k}_{\rm out} - \bar{k}_{\rm cen})$\\
&\\
\hline
\end{tabular}
\end{table}

\subsection{Initial conditions for emissivity and field ordering}
\label{em-field}

For a given velocity field in the jet, the adiabatic models require the
radiating particle density and magnetic field components to evolve
self-consistently with the velocity.  These models can therefore be
specified completely by setting their initial values at one point on any
given streamline, most straightforwardly at a surface of constant distance
from the nucleus $\rho = r = \bar{r}$.  We need to define the emissivity
variation $\epsilon(\bar{r},s)$ and two field-component ratios:
$j(\bar{r},s) = \langle B_r^2 \rangle^{1/2} /\langle B_t^2 \rangle^{1/2}$
(radial/ toroidal) and $k(\bar{r},s) = \langle B_l^2 \rangle^{1/2}
/\langle B_t^2 \rangle^{1/2}$ (longitudinal/toroidal). The number of free
parameters needed to specify an adiabatic model is much smaller than that
for the equivalent free model, in which the emissivity and field ordering
are allowed to vary separately as smooth functions of position.

We have made two sets of adiabatic models, attempting to fit different
regions of the jet, as follows:
\begin{enumerate}
\item the outer region alone, with initial conditions set at its
boundary with the flaring region ($\bar{r} = r_0$);
\item the flaring and outer regions, with initial conditions set at
the boundary between the inner and flaring regions ($\bar{r} = r_1$).
\end{enumerate}
The functional forms for the initial transverse variations of emissivity,
$\epsilon(\bar{r},s)$ and the field component ratios $j(\bar{r},s)$ and
$k(\bar{r},s)$ are given in Table~\ref{Trans-funcs}.

\subsection{Model integration, fitting and optimization}

The integration through the model jets to determine the $I$, $Q$ and $U$
brightness distributions is identical to that described by LB, except
for the determination of emissivity and field ordering at a given point
in the jet, the main steps in which are:
\begin{enumerate}
\item Determine coordinates in a frame fixed in the jet, in particular the
distance coordinate $\rho$ and the streamline index $s$, numerically if
necessary.
\item Evaluate the velocity at that point, together with the
angle between the flow direction and the line of sight $\psi$.  Derive
the Doppler factor $D = [\Gamma (1 - \beta\cos \psi)]^{-1}$ and hence the
rotation due to aberration ($\sin\psi^\prime = D\sin\psi$, where
$\psi^\prime$ is measured in the rest frame of the jet material).
\item Look up the initial values for the emissivity and field-ordering
parameters on the streamline.
\item Calculate the evolution of the particle density and field components
along the streamline using equations~\ref{n-eqn}, \ref{a-eqn}--\ref{vol-eqn}
and \ref{eq-dzs}--\ref{eq-shearfactor}.
\item Evaluate the emissivity function $\epsilon(\rho, s) \propto n
B^{1+\alpha}$.
\item Evaluate the position angle of polarization, and the rms field
components along the major and minor axes of the probability density
function of the field projected on the plane of the sky \citep{L02}.
Multiply by $\epsilon(\rho,s)D^{2+\alpha}$, to scale the emissivity and
account for Doppler beaming.
\item Derive the total and polarized emissivities using the expressions
given by \citet{L02} and convert
to observed Stokes $Q$ and $U$.
\end{enumerate}

Fitting to observed images and optimization of the models are done by
$\chi^2$ minimization as described in LB.

\begin{figure*}
\epsfxsize=17cm
\epsffile{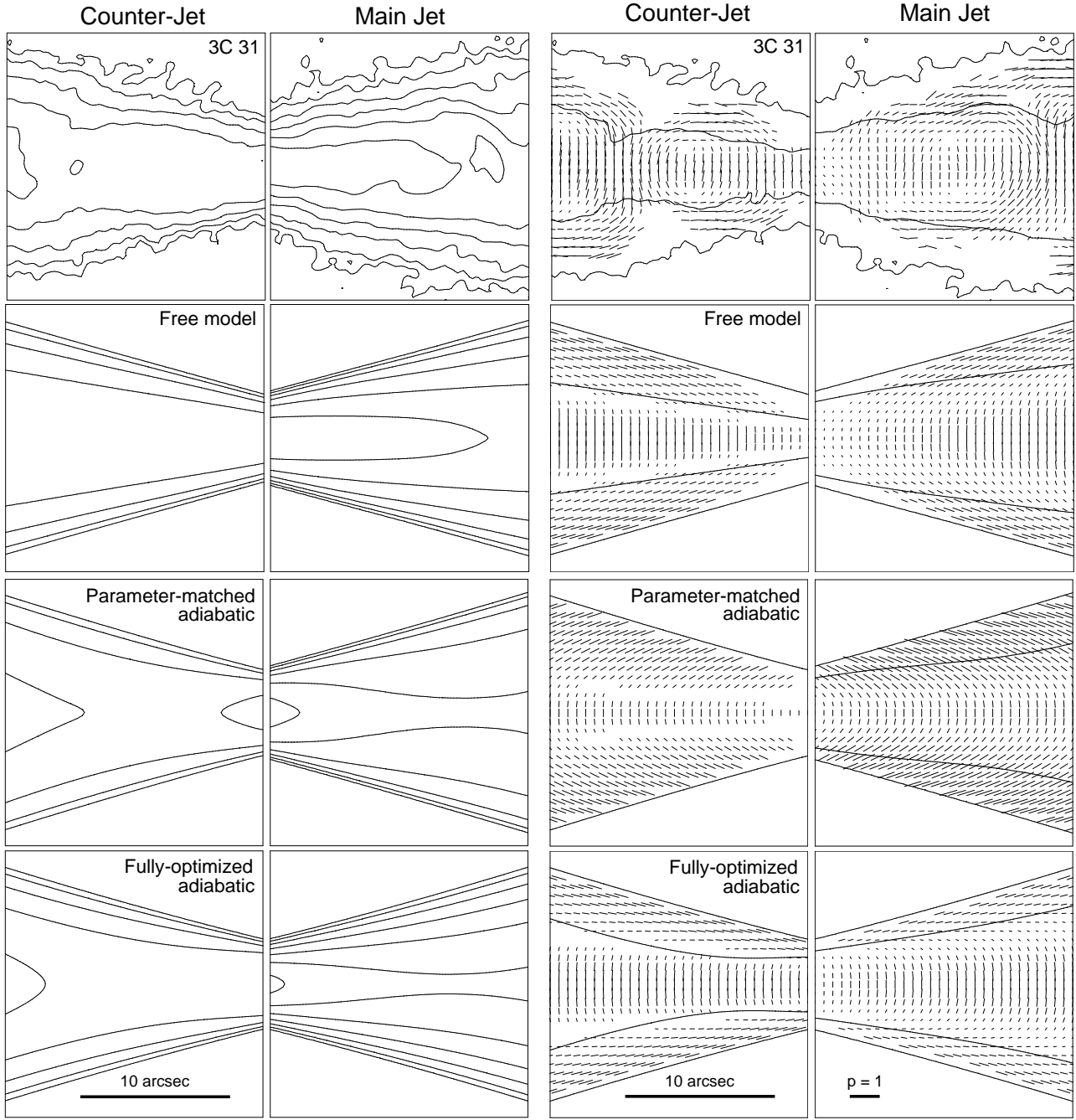}
\caption{Contours of total intensity and magnetic-field vectors at a resolution
of 0.75 arcsec for the outer regions of the jet and counter-jet.  The plots
cover the range 10 -- 27\,arcsec on either side of the nucleus and are in pairs
with the counter-jet on the left and the main jet on the right.  The angular
scale is indicated by the labelled bars at the bottom of the diagram. From the
top: observations; free SSL model; parameter-matched adiabatic model;
fully-optimized adiabatic model.  The initial conditions for the adiabatic
models are set at the boundary between the flaring and outer regions. The
left-hand pairs of panels show total intensity contours at levels of $-$1, 1, 2,
4, 8, 16, 32 $\times$ 20\,$\mu$Jy/beam area.  The right-hand pairs show vectors
whose magnitudes are proportional to $p$ and whose directions are those of the
apparent magnetic field, superimposed on selected $I$ contours. The scale of $p$
is indicated by the bar in the bottom right-hand panel.
\label{Out-contours}}
\end{figure*}

\section{Fits to the outer region alone}
\label{Outer-region}

We know (LB, Fig.\,20) that the analytical adiabatic approximation of
\citet {Bau97} is within a factor of two of the variation of emissivity
along the outer jet if the velocity profile is as estimated in our
best-fitting free models.  In this section, we therefore fit full
adiabatic models to the outer region alone, setting the initial conditions
at its boundary with the flaring region and and computing the $\chi^2$
value for the fit at distances $>$10\,arcsec from the nucleus, where all
lines of sight pass only through the outer region. We initially allow
the angle to the line of sight, the velocity field and the initial
emissivity and field-ordering parameters to vary. The results for the
Gaussian and SSL adiabatic models are very similar, but the former always
fit slightly better. As the Gaussian models also have fewer free
parameters, we concentrate on them when comparing two classes of adiabatic
models with the data, as follows:

\begin{enumerate}
\item A model having precisely the same angle to the line of sight, velocity
field, initial conditions and flux scaling as the best-fitting free
Gaussian model from LB, which it is therefore forced to match at the end
of the flaring region. We henceforth refer to this as the {\em
parameter-matched} adiabatic model. It has no free parameters.
\item A {\em fully-optimized} adiabatic model. The model flux density is
constrained to be the measured value for the outer region, but all other
parameters are varied to achieve the best fit.
\end{enumerate}
For the best-fitting free model, there is a well-constrained solution for
which the reduced $\chi^2$ in the outer region, $\chi^2_{\rm red}$ = 1.6
(LB).  For the adiabatic models, the optimization routine failed to find a
well-defined $\chi^2$ minimum: there is a broad range of solutions with
very similar brightness and polarization distributions, but different
angles to the line of sight.  We therefore fixed $\theta$ at values
separated by 5$^\circ$ in the range $20^\circ \leq \theta \leq 75^\circ$
and optimized the remaining parameters.  The minimum $\chi^2_{\rm red}$ =
1.9 for the Gaussian model occurs for $\theta \approx 55^\circ$, but there
are solutions with $\chi^2_{\rm red} < 2.5$ for $25^\circ \leq \theta \leq
70^\circ$.  The best-fitting values are listed in Table~\ref{Out-table},
along with those of the parameter-matched model. We have not performed an
error analysis because of the wide range of acceptable parameters.

\pagebreak

\begin{figure}
\epsfxsize=8.5cm
\epsffile{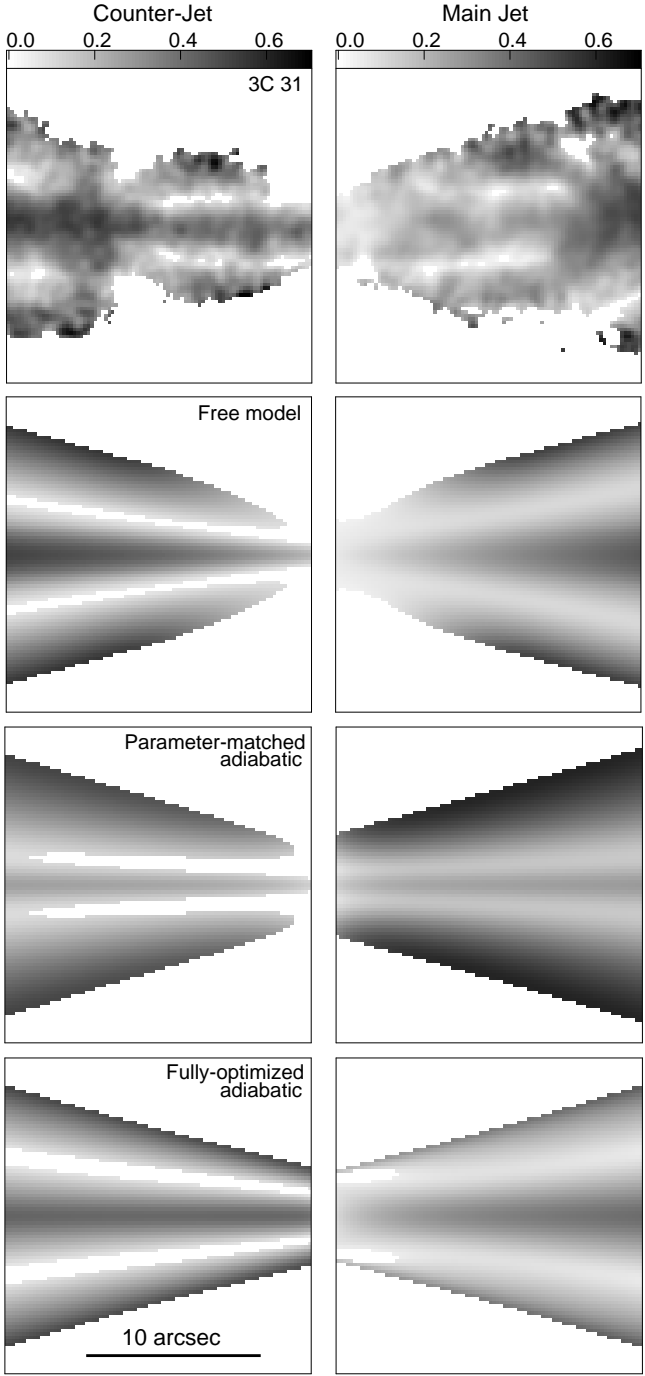}
\caption{Grey-scale representations of the degree of polarization, $p$, in the
outer jets.  The grey-scale runs from $p$ = 0 to $p$ = 0.7, as indicated by the
labelled wedges. The right- and left-hand sets of panels show the main and
counter-jets, respectively, and the angular scale is indicated by the labelled
bar at the bottom of the diagram. From the top: observations; free SSL model;
parameter-matched adiabatic model; fully-optimized adiabatic model.
\label{Out-pgrey}}
\end{figure}

\begin{figure}
\epsfxsize=8.5cm
\epsffile{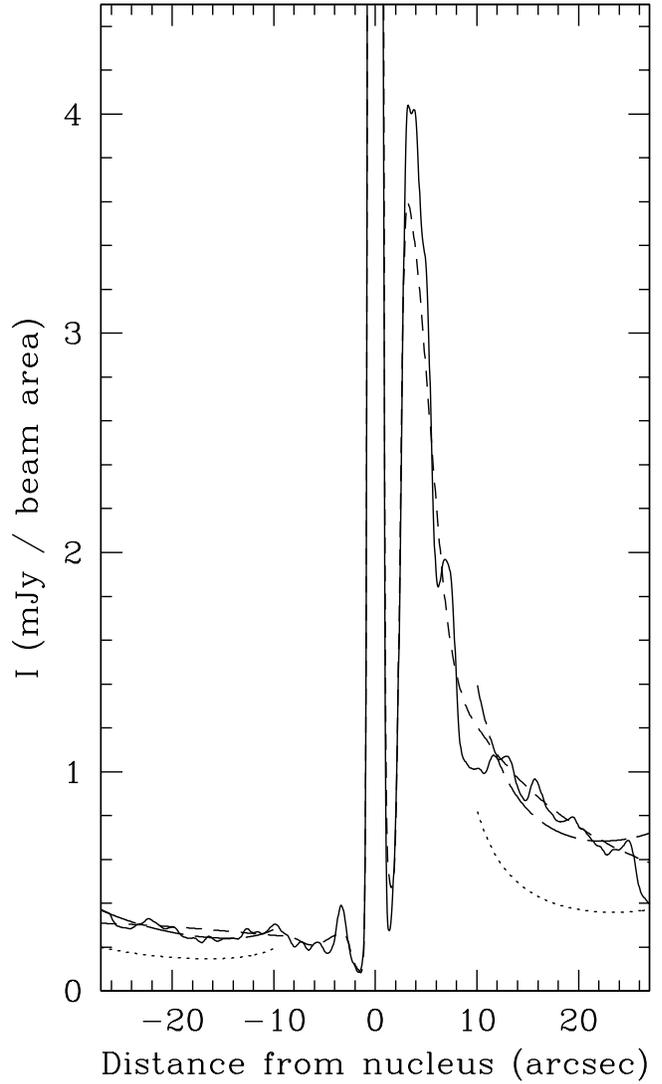}
\caption{Profiles of total intensity along the jet axes at a resolution of
  0.75\,arcsec FWHM. Full line: observations; short dashed line: free SSL model;
  dotted line: parameter-matched adiabatic model; long-dashed line: optimized
  adiabatic model.  The initial conditions for the adiabatic models are set at
  the end of the flaring region, so their profiles are plotted only beyond
  10\,arcsec from the nucleus, where the line of sight is entirely within the
  outer region.\label{Iout-profile}}
\end{figure}

\clearpage

Fig.~\ref{Out-contours} compares the data and the free and adiabatic
models by showing contours of total intensity for the outer regions of the
jets, the best-fitting free SSL model and the two Gaussian adiabatic
models.  It also plots vectors whose magnitudes are proportional to the
degree of polarization, $p$, and whose directions are those of the
apparent magnetic field.  Fig.~\ref{Out-pgrey} shows $p$ for the same
regions in a grey-scale representation and Fig~\ref{Iout-profile} shows
longitudinal profiles of total intensity.

\begin{table}
\caption{Parameters for adiabatic models of the outer
region.  The initial values are defined at the
boundary between the outer and flaring regions, $\bar{r} = r_0$. \label{Out-table}}
\begin{tabular}{lrr}
\hline
&&\\
Parameter &\multicolumn{1}{c}{Parameter-}&
\multicolumn{1}{c}{Fully-}\\
&\multicolumn{1}{c}{matched}&\multicolumn{1}{c}{optimized}\\
&&\\
\multicolumn{3}{c}{Geometry}\\
Angle to line of sight $\theta$ & 51.4 & 55.0 \\
&&\\
\multicolumn{3}{c}{Velocity field}\\
On-axis velocity $\beta_0$ & 0.54 & 0.76 \\
On-axis velocity $\beta_f$ & 0.27 & 0.38 \\
Fractional edge velocity $v_0$& 0.63 & 0.26 \\
&&\\
\multicolumn{3}{c}{Emissivity profile}\\
Fractional edge emissivity $\bar{e}$  & 0.26 & 0.26 \\
&&\\
\multicolumn{3}{c}{Radial/toroidal field ratios}\\
On-axis $\bar{j}_{\rm cen}$ & 0.0  & 0.53 \\
Edge    $\bar{j}_{\rm out}$ & 0.92 & 0.0  \\
Index   $p$           & 0.41 & 1.09 \\
\multicolumn{3}{c}{Longitudinal/toroidal field ratios}\\
On-axis $\bar{k}_{\rm cen}$ & 0.82 & 2.53 \\
Edge    $\bar{k}_{\rm out}$ & 0.82 & 0.0  \\
Index   $q$           & 0.0  & 0.55 \\
&&\\
\multicolumn{3}{c}{Goodness of fit}\\
$\chi^2_{\rm red}$    & 8.64 & 1.89 \\
&&\\
\hline
\end{tabular}
\end{table}

The main features evident from this comparison are as follows:
\begin{enumerate}
\item The total intensity predicted by the parameter-matched adiabatic
model falls off rapidly close to the boundary between the flaring and
outer regions (where it is forced to match the free model) and thereafter
is much lower than the observed values.  This is most clearly shown by the
longitudinal profile in Fig.~\ref{Iout-profile} and is reflected in the
high $\chi^2_{\rm red} = 8.6$.
\item The total-intensity distribution for the fully-optimized adiabatic model
is in much better agreement with the observations. Some of the improvement
in $\chi^2$ results from the requirement to fit the total flux density in
the outer region rather than to match the free model exactly at the start
of the region. The model parameters can then be optimized to give a
good fit between 11 and 25\,arcsec from the nucleus, at the expense of an
error between 10 and 11\,arcsec, where the initial decrease of brightness
with distance from the nucleus is slightly too rapid, and at distances
\ga\ 25\,arcsec, where the model intensity is too high
(Figs~\ref{Out-contours} and \ref{Iout-profile}). The total intensity of
the counter-jet is well fitted by this model.
\item Both adiabatic models predict too high a degree of polarization in
the transverse-field region on the axis of the main jet, especially within
$\approx$10\,arcsec of the start of the outer region
(Fig.~\ref{Out-pgrey}).
\item Conversely, both adiabatic models give slightly too low a degree of
polarization on the axis of the counter-jet (Figs.~\ref{Out-contours} and
\ref{Out-pgrey}).
\item The degree of polarization at the edges of both jets is
overestimated, again within $\approx$10\,arcsec of the start of the outer
region. (Figs.~\ref{Out-contours} and \ref{Out-pgrey})
\item The field-vector directions for the fully-optimized adiabatic model
are close to those observed, but those for the parameter-matched model are
incorrect at the edge of the jet (where they should be parallel to the
surface; Fig.~\ref{Out-contours}).
\end{enumerate}

The problems encountered in fitting the polarization of the outer region with
these laminar-flow adiabatic models are fundamental and do not depend on the
choice of $\theta$ or the velocity field.  The free models fitted by LB
introduce a significant radial magnetic field component, $B_r$, at the edge of
the jet at the start of the outer region to produce the low degree of
polarization $p$ there.  This field component also reduces the values of $p$
on-axis. The very high degrees of polarization observed at the jet edges further
from the nucleus require $B_r$ again to become small compared with the toroidal
and longitudinal components $B_t$ and $B_l$ there.  This cannot be achieved in
our adiabatic models, for which the ratio $B_r/B_t$ is fixed by our choice of an
axisymmetric velocity field with straight streamlines for the outer region (the
coefficients of ${\bf \hat{a}}$ and ${\bf \hat{b}}$ in equations~\ref{a-eqn-lin}
and \ref{b-eqn-lin} are identical).  More generally, $B_r/B_t$ alters very
slowly unless the rate of change of expansion is large: this is a direct
consequence of flux-freezing in a laminar velocity field (equations~\ref{a-eqn}
and \ref{b-eqn}).  In order to change the ratio, it is necessary to introduce a
component of velocity across the existing model streamlines. This component
could result from turbulence, for example as a result of entrainment of the
external medium. The reason why $B_r$ disappears in the outer parts of the jets
remains unclear.

As a result of the small range of jet/counter-jet sidedness ratio in the
outer region, the fully-optimized adiabatic models are poorly
constrained. This results in a degeneracy between the velocity and angle
to the line of sight: the optimized values of the fiducial velocities
$\beta_0$ and $\beta_f$ range from 0.33 and 0.16 for $\theta$ = 25$^\circ$
to 0.87 and 0.43 for $\theta$ = 70$^\circ$.  For the best-fitting model,
the velocities are 0.71 and 0.34 (Table~\ref{Out-table}). These are
significantly larger than the velocities deduced for the free model (0.54
and 0.27) and imply faster deceleration. In addition, the fractional
velocity at the edge of the jet is lower (0.26, compared with 0.63). The
combination of these two differences leads to much higher shear at the
edges of the jet for the optimized model. The changes result from a
partially-successful attempt to fit the slow brightness decline at the
start of the outer region and the apparent magnetic-field direction at the
edge of the jet.  In effect, the coupling that the adiabatic models
require between the variations of the field geometry and the emissivity
forces them to steeper velocity gradients in the outer region than those
in the best-fitting free model. We have also investigated the effects of
changing the {\em functional form} of the velocity law as well as the
parameters. It is possible to improve the fit to the total intensity by
making the velocity decrease more rapidly at the start of the outer region
and more slowly at large distances, but this always gives a worse fit to
the polarization, and we have been unable to improve the overall $\chi^2$.

\begin{figure*}
\epsfxsize=17cm
\epsffile{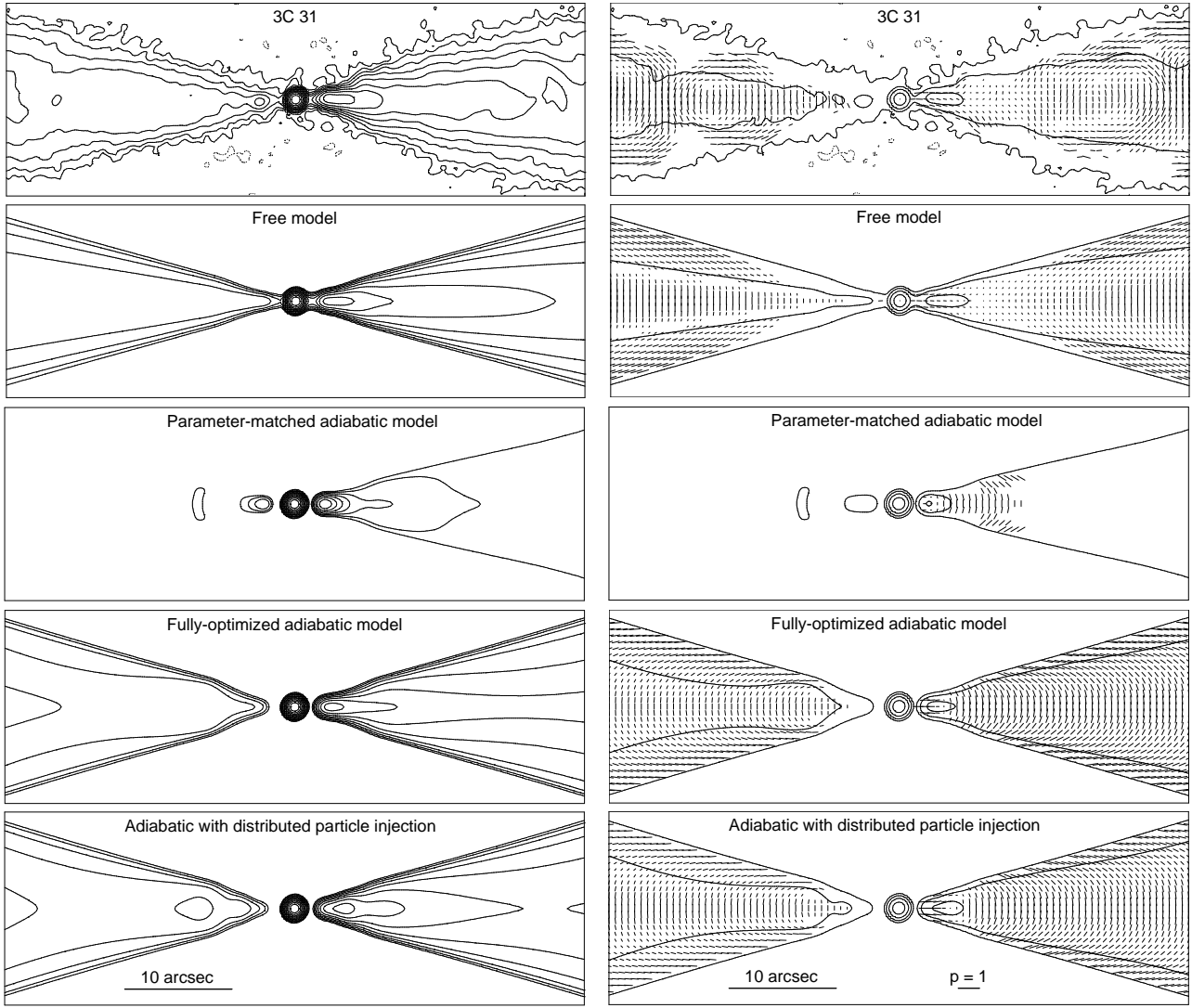}
\caption{Contours of total intensity and magnetic-field vectors at a
resolution of 0.75 arcsec. The plots cover 27\,arcsec on either side of
the nucleus and the angular scale is indicated by the labelled bars. From
the top: observations; free SSL model; parameter-matched adiabatic model;
fully-optimized adiabatic model; adiabatic model with distributed particle
injection.  The initial
conditions for both adiabatic models are set at the boundary between the
inner and flaring regions and the
inner region is not modelled. The
left-hand panels show total intensity contours at levels of $-$1, 1, 2, 4,
8, 16, 32 $\times$ 20\,$\mu$Jy/beam area.  The right-hand pairs show
vectors whose magnitudes are proportional to $p$ and whose directions are
those of the apparent magnetic field, superimposed on selected $I$
contours. The scale of $p$ is indicated by the bar in the bottom
right-hand panel. No emission is calculated in the inner region
($\pm$2.5\,arcsec) for the adiabatic models.
\label{Ivec}}
\end{figure*}

The parameter-matched model effectively incorporates information gained
from free-model fits to the inner jet regions and should therefore provide
a much more realistic description of the velocity field, but its predicted
brightness distribution initially declines too rapidly with distance from
the nucleus (Figs~\ref{Out-contours} and \ref{Iout-profile}).  Particle
injection and/or field amplification might occur close to the boundary
with the flaring region, thereby slowing the emissivity decline
(cf. Section~\ref{Inject}).

\section{Fits that include the flaring region}
\label{Whole-jet}

\subsection{The inner region}
\label{Inner-jet}

As noted by LB (Section\,5.4), the conical inner region
(the first 2.5 arcsec of the jet) shows no evidence
for deceleration, but neither does it exhibit the extremely rapid brightness
fall-off characteristic of adiabatic expansion at constant speed. The low
sidedness ratio observed in this region also led us to suggest that its
emission comes mostly from a slow surface layer which does not persist to
larger distances. We have insufficient resolution to build or verify a
model of the transverse structure of this region, so we do not consider
it further in this paper.

\subsection{Models with initial conditions set at the boundary between
the inner and flaring regions}

LB showed that the analytical adiabatic approximation of \citet{Bau97} fails
completely for the flaring region. We now examine whether this conclusion would
be modified by including more realistic initial conditions and the effects of
velocity shear by modelling the flaring and outer regions of 3C\,31, setting the
initial emissivity and field-ordering profiles at the boundary between the inner
and flaring regions ($\bar{r} = r_1$).  As in Section~\ref{Outer-region}, we
find that the results for Gaussian and SSL velocity profiles are very similar,
but that the former fit better, as well as having fewer free parameters.  We
therefore again show only the Gaussian-velocity case, and compare
fully-optimized models with a parameter-matched model whose initial conditions,
velocity field, angle to the line of sight and flux scaling are identical to
those of the best-fitting free model.

The best fits for the optimized models are much poorer ($\chi^2_{\rm red}
\approx 5$) than those for the free models ($\chi^2_{\rm red}$ = 1.7 -- 1.8),
and also require extremely high velocities at the start of the flaring
region. As for the outer region alone (Section~\ref{Outer-region}), solutions of
comparable quality can be found over a wide range of angles to the line of
sight ($30^\circ \leq \theta \leq 55^\circ$). We show the best-fitting example,
which again has $\theta = 55^\circ$.

\begin{figure}
\epsfxsize=8.5cm
\epsffile{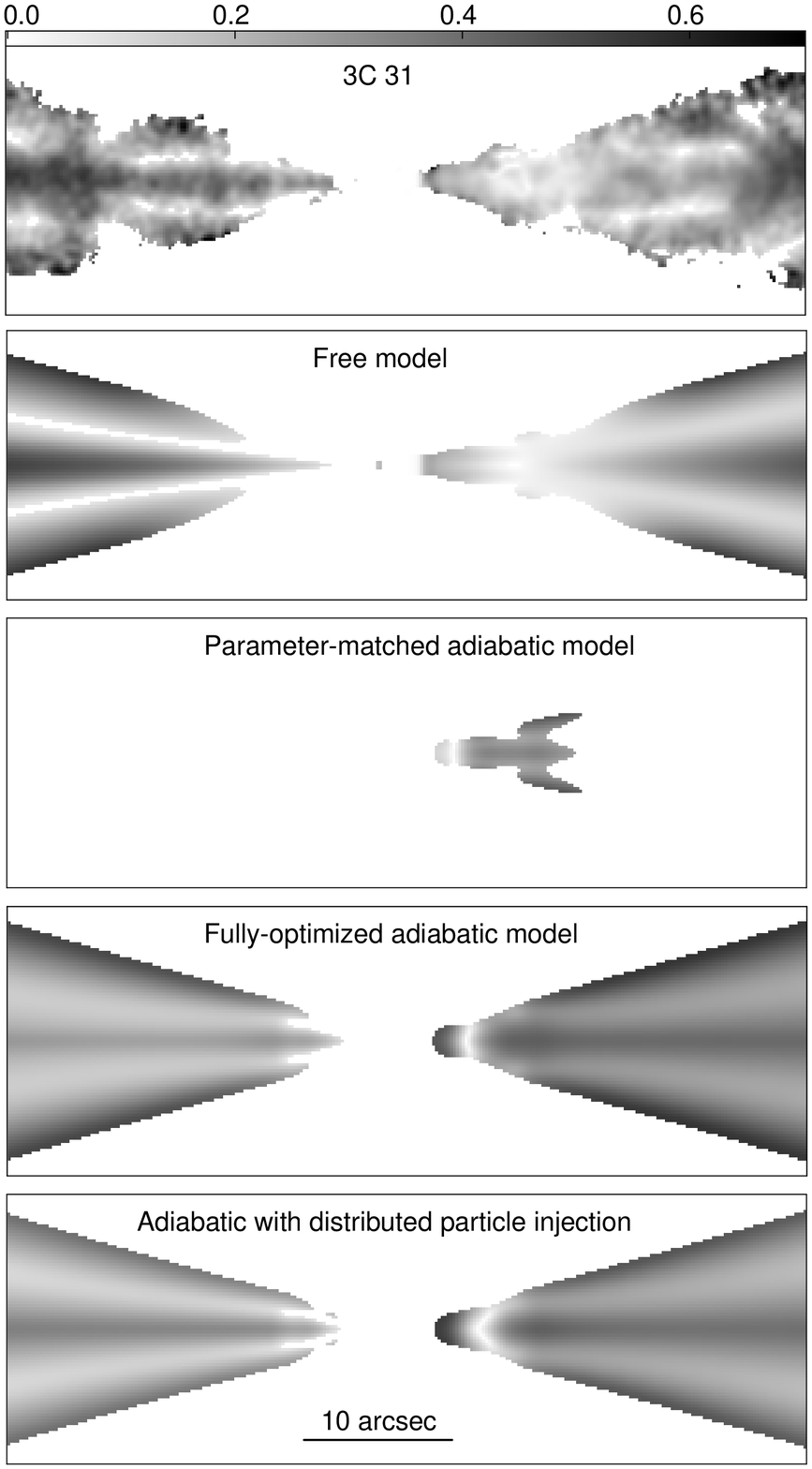}
\caption{Grey-scale representations of the degree of polarization.  The
grey-scale runs from $p$ = 0 to $p$ = 0.7, as indicated by the labelled
wedge. The panels extend $\pm$27\,arcsec from the nucleus and the angular
scale is indicated by the labelled bar. From the top: observations; free
SSL model; parameter-matched adiabatic model; fully-optimized adiabatic
model; adiabatic model with distributed particle injection. The initial
conditions for the adiabatic models are set at the boundary between the
inner and flaring regions. No emission is calculated for the adiabatic
models within 2.5\,arcsec of the nucleus.\label{Pgrey}}
\end{figure}

\begin{figure}
\epsfxsize=8.5cm
\epsffile{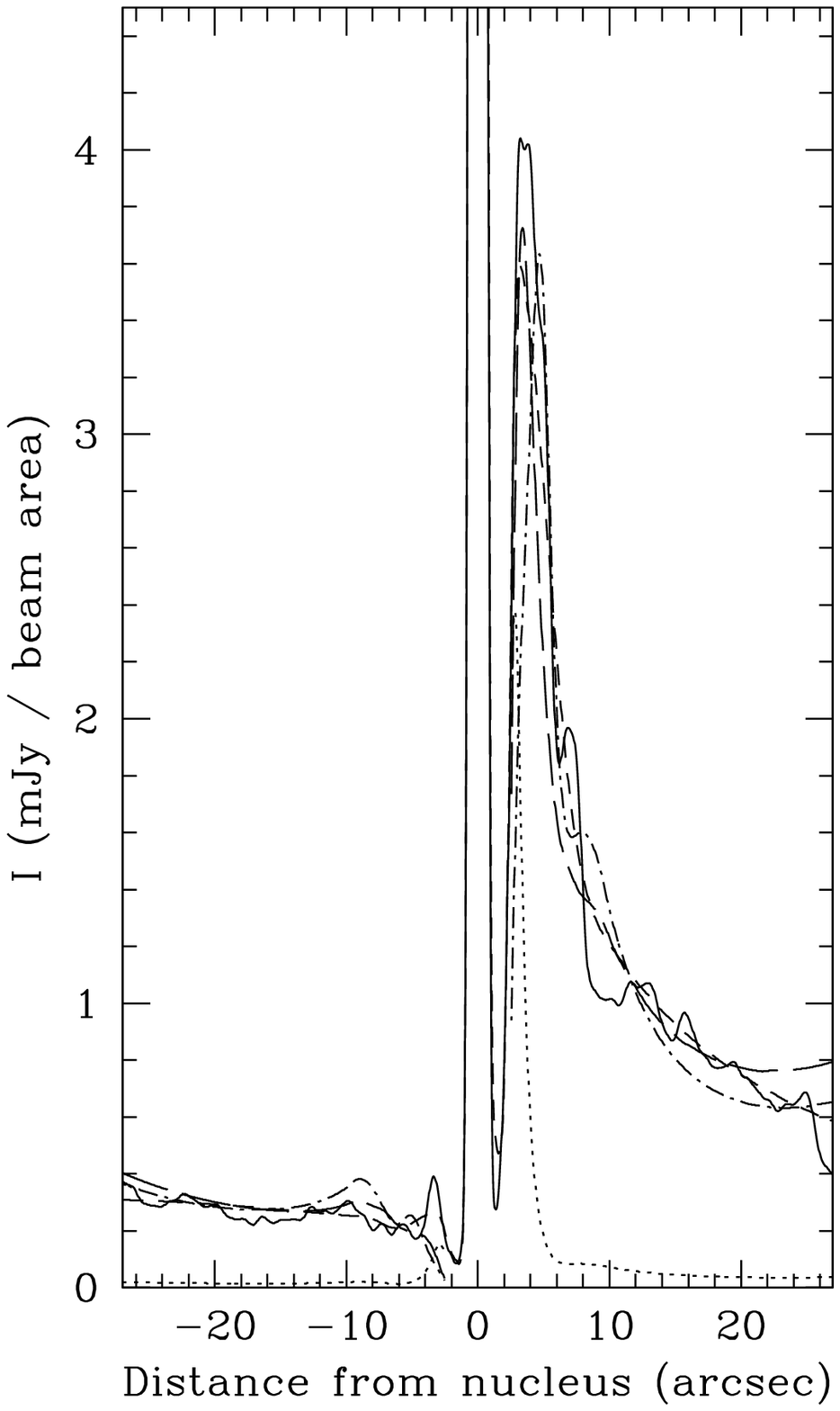}
\caption{Profiles of total intensity along the jet axes at a resolution of
  0.75\,arcsec FWHM. Full line: observations; short dashes: free SSL model;
  dots: parameter-matched adiabatic model; long dashes: fully-optimized
  adiabatic model; dash-dots: adiabatic model with distributed particle
  injection. The initial conditions for the adiabatic models are set at the
  boundary between the inner and flaring regions, so their profiles are plotted
  only for the outer and flaring regions ($>$2.5\,arcsec from the
  nucleus).\label{Iprofile}}
\end{figure}

\begin{figure}
\epsfxsize=7.5cm
\epsffile{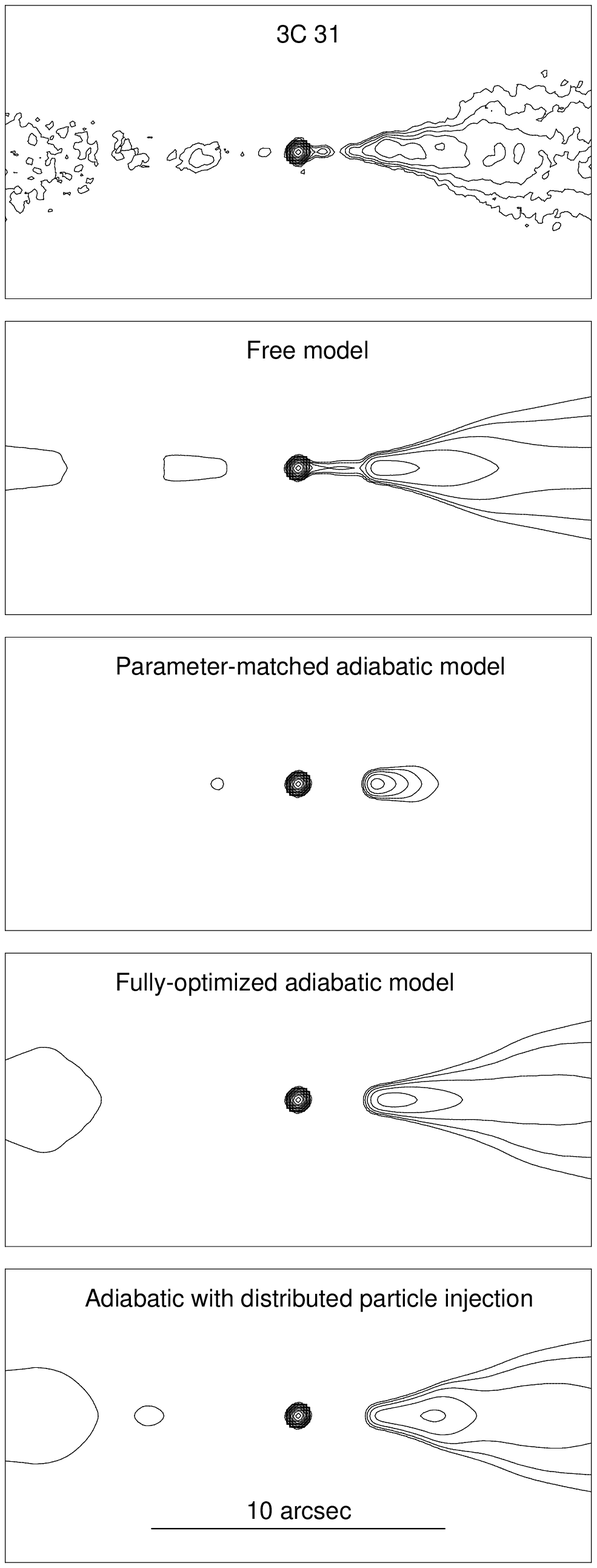}
\caption{Contours of total intensity at a resolution of 0.25\,arcsec FWHM.
The panels cover $\pm$10\,arcsec from the nucleus and the angular scale is
indicated. The contour levels are $-$1, 1, 2, 4, 8, 16, 32 $\times$
30\,$\mu$Jy/beam area. From the top: observations; free SSL model;
parameter-matched adiabatic model; fully-optimized adiabatic model;
adiabatic model with distributed particle injection.  As the adiabatic
models have initial conditions set at the boundary between the inner and
flaring regions, no emission is calculated for the inner jet region
($<$2.5\,arcsec from the nucleus).
\label{Ihires}}
\end{figure}

Images of total intensity and linear polarization for the parameter-matched and
fully-optimized models are compared with the observations of 3C\,31 and the
best-fitting free SSL model in Figs~\ref{Ivec} -- \ref{Ihires} and the
parameters for both models are listed in Table~\ref{Params}.  The main points of
interest are as follows.
\begin{enumerate}
\item The parameter-matched model fails completely to fit the brightness
distribution: its initial brightness fall-off is far too steep on both
sides of the nucleus. This confirms our conclusion from the analytical
approximation.
\item The polarization distribution predicted by the parameter-matched
adiabatic model is also incorrect, the apparent magnetic field being
almost perpendicular to the jets at their edges, rather than parallel as
observed.
\item The fully-optimized model fits the total intensity from the main jet
and the outer counter-jet fairly well (Figs~\ref{Ivec}, \ref{Iprofile} and
\ref{Ihires}), but seriously underestimates the emission from the
counter-jet close to the beginning of the flaring region, where the
predicted jet/counter-jet sidedness ratio is $\approx$50, compared with the
observed maximum of 13 for the entire region.
\item The fully-optimized model shows a qualitatively
correct polarization distribution, with transverse apparent field on the
axis of both jets and longitudinal field at the edges. It predicts a
higher degree of polarization on-axis in the main jet than in the
counter-jet (opposite to the observed difference) and also fails to
reproduce the regions of low polarization at the edges of both jets and
across the whole of the main jet in the flaring region (Figs~\ref{Ivec}
and \ref{Pgrey}).
\end{enumerate}

The parameter-matched model fails because there is insufficient deceleration to
counteract adiabatic losses resulting from the rapid expansion of the jets in
the flaring region, even when field amplification by velocity shear is taken
into account. There is also insufficient shear to counteract the effects of
expansion at the edges of the jets, leading to a transverse apparent field
there.  The fully-optimized model fits much better, but has two main
problems. First, it requires an extremely rapid deceleration in the flaring
region: the initial on-axis velocity $\beta_1$ is in the range 0.93 -- 0.99 for
any $\theta$ and the corresponding velocity at the edge of the jet is \ga 0.7,
so high sidedness ratios are inevitable.  Second, as in the outer region, the
polarization data require far larger a change in the ratio $B_r/B_t$ than is
allowed in our axisymmetric, laminar adiabatic model: a significant radial field
component must first be generated at the edge of the flaring region and then
destroyed further from the nucleus.

The extremely high sidedness ratios required by the optimized adiabatic
models are inconsistent with observations. To help adiabatic models fit
the data using the more realistic velocity field of the free models,
additional emissivity must be introduced in the flaring region, as we now
discuss.

\begin{table}
\caption{Fitted parameters for models with initial conditions at the
boundary between the inner and flaring regions ($\bar{r} = r_1$). Note that
$\theta$ and the velocity field are fixed for the particle-injection
model; its remaining parameters are optimized. \label{Params}}
\begin{minipage}{120mm}
\begin{tabular}{lrrr}
\hline
&&&\\
Quantity &\multicolumn{1}{c}{Parameter-}
&\multicolumn{1}{c}{Fully-}
&\multicolumn{1}{c}{Particle}\\
&\multicolumn{1}{c}{matched}
&\multicolumn{1}{c}{optimized}
&\multicolumn{1}{c}{injection}\\
&&&\\
\hline
&&&\\
\multicolumn{4}{c}{Angle to line of sight}\\
 $\theta$ (degrees) & 51.4 & 55.0 & 51.4 \\
&&&\\
\multicolumn{4}{c}{Velocity field}\\
On-axis velocities &&&\\
$\beta_1$ & 0.76 & 0.99   & 0.76 \\
$\beta_0$ & 0.54 & 0.77  & 0.54 \\
$\beta_f$ & 0.27 & 0.40  & 0.27 \\
Edge velocities &&&\\
$v_1$     & 0.97 & 0.93  & 0.97 \\
$v_0$     & 0.63 & 0.30  & 0.63 \\
velocity exponent $H$     & 8.82 & 2.00   & 8.82 \\
&&&\\
\multicolumn{4}{c}{Emissivity profile}\\
Edge emissivity $\bar{e}$ & 0.37 & 0.05  & 0.18  \\
&&&\\
\multicolumn{4}{c}{Radial/toroidal field ratios}\\
On-axis $\bar{j}_{\rm cen}$ & 0.0 & 0.76   & 0.81  \\
Edge  $\bar{j}_{\rm out}$ &  0.78 & 0.42  & 0.42  \\
Index $p$ & 0.41 & 1.50  & 0.62  \\
&&&\\
\multicolumn{4}{c}{Longitudinal/toroidal field ratios}\\
On-axis $\bar{k}_{\rm cen}$ &  1.17 & 4.97  & 3.01  \\
Edge  $\bar{k}_{\rm out}$ & 1.17   & 2.65  &  3.01\\
Index $q$ & 0.0 & 3.96  & 0.0  \\
&&&\\
\multicolumn{4}{c}{Particle injection parameters}\\
Cut-off $\rho_{\rm inj}$ (kpc) &&& 2.16 \\
Exponent $Q$ &&& -4.71\\
Normalization $\Delta_0$ &&& 84.5\\
&&&\\
\multicolumn{4}{c}{Goodness of fit}\\
$\chi^2_{\rm red}$ & 30.4  & 4.65  & 4.92 \\
&\\
\hline
\end{tabular}
\end{minipage}
\end{table}

\subsection{Particle injection}
\label{Inject}

It is clear from the earlier discussion that processes other than adiabatic
evolution in an axisymmetric, laminar velocity field must be important in the
flaring region of 3C\,31 and may also have some effect at the start of the outer
region.  A process that locally increases the emissivity is clearly needed to
compensate adiabatic expansion losses in the flaring region. Our modelling
approach lets us estimate the rest-frame emissivity and the field-component
ratios, but it cannot disentangle the roles of particles and magnetic field in
the absence of any constraint from inverse-Compton emission. The observation of
X-ray emission from the inner and flaring regions shows that fresh relativistic
particles must be injected there, as the synchrotron lifetimes of electrons
radiating at these frequencies are $\sim$10's of years \citep{Hard02}.  We
therefore investigate a simple, but self-consistent physical model in which new
(e.g. reaccelerated) particles with an energy distribution $\Delta n(E)dE
\propto E^{-(2\alpha+1)}dE$ are added in a distributed fashion over the flaring
region and then evolve adiabatically. We again fix the angle to the line of
sight and velocity field at the best-fitting free model values, and assume that
the magnetic field is frozen into the flow. We then optimize the initial
conditions and particle injection function.  We parameterize the particle
injection as the ratio of the additional emissivity at position $(\rho, s)$ in
the flaring region to its value on the same streamline at the start of the
region, using the empirical relation:
\[ \frac{\Delta\epsilon(\rho,s)}{\epsilon(r_1,s)} = \Delta_0 \left [ 1 +
\left (\frac{\rho-r_1}{r_0-r_1} \right )\right ]^{-Q} \] for
$r_1 \leq \rho \leq \rho_{\rm inj}$ (the emissivity function $\epsilon$
is defined in Section~\ref{assumptions}).  This means that the transverse
injection profile is a constant Gaussian function. The parameters
$\Delta_0$, $Q$ and $\rho_{\rm inj}$ are optimized, with $\rho_{\rm
inj}$ restricted to $r_1 \leq \rho_{\rm inj} \leq r_0$.

\begin{figure}
\epsfxsize=8.3cm
\epsffile{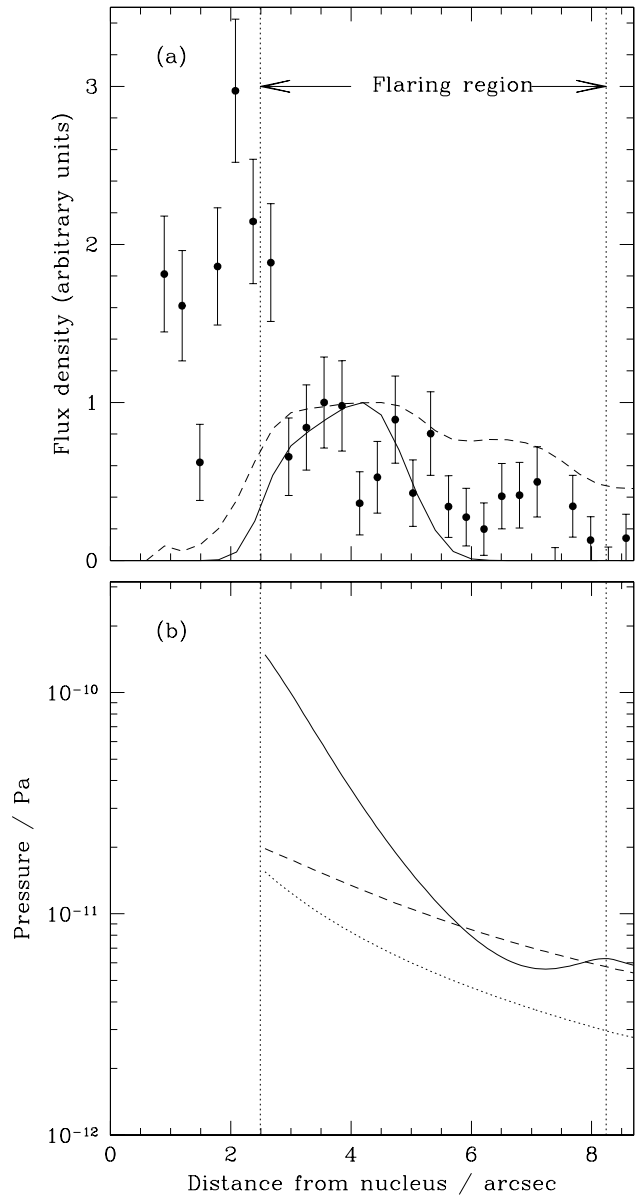}
\caption{Profiles of X-ray and radio emission and pressure along the axis
of the main jet.  (a) Flux-density profiles (on arbitrary scales) derived
as in Fig.\,3 of \citet{Hard02} by integration in boxes of size 0.3
$\times$ 2\,arcsec$^2$ (short axis along the jet) on images of resolution
0.6\,arcsec FWHM. Points: observed X-ray emission \citep{Hard02}; dashed
line: observed radio emission (LB); full line: predicted X-ray emission
profile for the flaring region alone, as described in the text.  Note that the
radio profile is derived from an image at a lower resolution than that used to
make the equivalent plot in \citet{Hard02} in order to match the {\em Chandra}
point-spread function as closely as possible.
(b)  Profiles of the internal (full), external (dashed) and
synchrotron minimum (dotted) pressures from our conservation-law analysis
(\citealt{LB02b}, Fig. 4, but projected onto the plane of the sky). The vertical
dotted lines mark the extent of the flaring region as defined by LB and in
Section~\ref{Geom}.
\label{Xray-inject}}
\end{figure}

The fit to the total-intensity distribution is generally good (Figs~\ref{Ivec},
\ref{Iprofile} and \ref{Ihires}), but the jet/counter-jet sidedness ratio close
to the boundary between the inner and flaring regions is still significantly
higher than is observed, despite the more modest velocity there. The reason is
that the field is almost purely longitudinal, so the ratio is $\approx[(1 +
\beta\cos\theta)/(1 - \beta\cos\theta)]^{3+2\alpha}$ \citep{Beg93} rather than
the usual $\approx[(1 + \beta\cos\theta)/(1 - \beta\cos\theta)]^{2+\alpha}$
appropriate for a field which is closer to isotropy.  Given that the assumed
magnetic-field evolution is unchanged, it is not surprising that the
deficiencies in the polarization fits noted earlier are still present
(Figs~\ref{Ivec} and \ref{Pgrey}) and that an unrealistic field structure is
required at the start of the flaring region. We do not, therefore, regard the
high sidedness ratio as a significant problem for the model.  The
fit to the total intensity in the outer region is very similar to that of the
parameter-matched model described in Section~\ref{Outer-region}, as expected, so
the need for some additional emissivity
there remains.
The overall fit is
slightly worse than for the fully-optimized model ($\chi^2_{\rm red} = 4.9$
compared with 4.7), but with 10 free parameters instead of 14. The optimized
parameters for the particle-injection model are given in Table~\ref{Params}. Key
implications are as follows:
\begin{enumerate}
\item Particle injection is required out to 6.4\,arcsec (2.2\,kpc)
from the nucleus, i.e.\ only over the first half of the flaring
region.  In the second half of the region, the jet decelerates
rapidly, thereby slowing the brightness decline predicted by adiabatic
models. Further particle injection there would produce too slow a
decline in brightness.
\item The injection rate decreases rapidly with distance from the
nucleus within that region.
\item The emissivity everywhere in the flaring region is dominated by
particles injected locally rather than those originating from the inner
region.
\end{enumerate}
Given the short synchrotron lifetimes of the electrons responsible for the
high-energy emission, X-rays are likely to be produced close to the injection
sites and the X-ray emissivity might therefore be expected to be roughly
proportional to the injection function, $\Delta$, rather than to the radio
emissivity.  A comparison between the observed X-ray and radio emission profiles
(\citealt{Hard02}, LB) and a model profile derived on this assumption
(Fig.~\ref{Xray-inject}) supports  this idea. The brightest
X-ray emission comes from the inner 2.5\,arcsec of the jet, which we have not
modelled, but for which non-adiabatic evolution is clearly required
(Section~\ref{Inner-jet}).  In the first half of the flaring region, where we
infer the need for significant particle injection, there is indeed evidence for
more X-ray emission than in the outer half of the region, where injection is not
required. The X-ray data in the flaring region are, however, too noisy to be
sure whether the ratio of X-ray to radio emission changes within this region as
our particle-injection model predicts. We also note that
high-resolution studies of X-ray and radio emission from jets have shown that
both may be highly inhomogeneous, with a complex mutual relationship
\citep{CenAX,M87X}.

If particles are indeed injected without much modifying the magnetic field, the
ratio of particle to field energy is likely to increase (by an amount which we
cannot estimate without a proper description of the injection mechanism). We
would not, therefore, expect equipartition between field and particle energy
everywhere in the flaring region, even if it holds in some locations.  The
region where we infer significant particle injection is also where the expansion
rate of the jet is increasing (Fig.~\ref{Geom-sketch}) and is where
\citet{LB02b} inferred a significant over-pressure (Fig.~\ref{Xray-inject}b).
The internal pressure significantly exceeds not only the external gas pressure,
but also the synchrotron minimum pressure, so quite large deviations from
equipartition could occur between 2.5 and $\approx$5\,arcsec from the nucleus.
In any case, we know that the magnetic field component ratios change in a way which
is inconsistent with flux freezing in a laminar velocity field, so it is very
likely that amplification of the field also occurs.

All of these results are consistent with injection (or reacceleration) of
radiating particles over a wide energy range in the first half of the flaring
region. The X-ray-emitting particles have extremely short synchrotron lifetimes
and radiate where they are injected, whereas the lower-energy, radio-emitting
particles suffer mainly adiabatic losses and radiate over the outer parts of the
jet, as we have calculated.  A quantitative study of the acceleration and
energy-loss processes over the entire energy spectrum is outside the scope of
the present paper.

\section{Summary and Conclusions}
\label{Conclusions}

We have investigated the hypothesis that the radio jets in 3C\,31 can be
modelled as adiabatic, decelerating, relativistic flows. Our technique has
several advantages over previous work in: modelling linear polarization as
well as total intensity; including the effects of velocity shear; taking
account of anisotropic emission in the rest frame and fitting to
two-dimensional images rather than to longitudinal profiles.

We have found that optimized adiabatic models give a fair description of the
observed brightness and polarization distributions in the outer parts of the
modelled region.  Their fit to the data is inferior to that of the free models
of LB, but is obtained with many fewer free parameters.  The adiabatic models
cannot describe the inner or flaring regions of the jets in 3C\,31, however: the
predicted distributions of total intensity and linear polarization are
inconsistent with those observed.  In the innermost region, the jets are clearly
non-adiabatic unless the emission comes primarily from a part of the jet volume
which is not expanding with distance from the nucleus (we do not resolve the
region transversely). In the flaring region, a much higher initial velocity is
required than is allowed by our measurements of jet/counter-jet sidedness
ratio. The emission in this region is well resolved, and comes from the entire
jet volume.

We have shown that a modified adiabatic model can still be fitted to the total
intensity in this region if we add distributed injection of relativistic
particles which then evolve adiabatically; the region where these particles must
be injected is also one where there is independent evidence for recent particle
acceleration from the detection of X-ray synchrotron radiation \citep{Hard02}
and of a local over-pressure in the jet from dynamical arguments \citep{LB02b}.

While the total intensity distributions in the jet and counter-jet of
3C\,31 can be reproduced satisfactorily by decelerating adiabatic jet
models with particle injection in the flaring region, the polarization
data (degree of linear polarization and apparent magnetic field direction)
cannot.  The apparent magnetic field configuration contains several
features that are qualitatively incompatible with adiabatic evolution of
all the field components in laminar-flow models, even in the presence of a
velocity shear.  We infer that the departures from adiabatic conditions in
3C\,31 must also include deviations either from the flux-freezing
description of the magnetic fields or from a laminar, axisymmetric
velocity field.  The free models of LB achieved better consistency with
the observed polarization structure by allowing the field to become
roughly isotropic at the edges of the jet in the flaring region.  This
could be achieved by a turbulent velocity component even if the field is
frozen into the flow, in which case the resulting shear would also
contribute to the enhancement in emissivity required to fit the adiabatic
models in the flaring region. Processes such as dynamo action or
field-line reconnection might also be significant.

More detailed analysis of the particle injection process inferred for the
flaring region would benefit from more sensitive X-ray, optical and infra-red
data and higher-resolution radio observations of this region, i.e.\ from deeper
exposures with {\em Chandra} and {\em HST} and from deep higher-resolution
polarimetry with the EVLA.  Improved knowledge of the intensity and apparent
magnetic field structures within this region might also assist development of
models for the magnetic field microphysics in this region of the 3C\,31 jet,
which all of our analysis suggests is crucial in determining the deceleration
dynamics of the jet, and its brightness and polarization properties further from
the nucleus.

\section*{Acknowledgments}

RAL would like to thank the National Radio Astronomy Observatory and Alan
and Mary Bridle for hospitality during this project.  The National Radio
Astronomy Observatory is a facility of the National Science Foundation
operated under cooperative agreement by Associated Universities, Inc.

\end{document}